\documentstyle[aps,prd,epsf]{revtex}

\newcommand{\be}{\begin{equation}}
\newcommand{\ee}{\end{equation}}
\newcommand{\ben}{\begin{eqnarray}
\displaystyle}
\newcommand{\een}{\end{eqnarray}}

\newcommand{\p}{\partial}
\newcommand{\na}{\nabla}

\newcommand{\tiA}{{\tilde A}}

\newcommand{\ep}{\epsilon}
\newcommand{\bep}{\bar \epsilon}

\newcommand{\bxi}{\bar \xi}
\newcommand{\balpha}{\bar \alpha}
\newcommand{\bz}{\bar z}
\newcommand{\bv}{\bar v}

\newcommand{\om}{\omega}
\newcommand{\ga}{\gamma}

\begin{document}

\title{Uniqueness Theorem for Static Dilaton $U(1)^N$ Black Holes}

\author{Marek Rogatko}

\address{Institute of Physics \protect \\
Maria Curie-Sklodowska University \protect \\
20-031 Lublin, pl.Marii Curie-Sklodowskiej 1, Poland \protect \\
rogat@tytan.umcs.lublin.pl \protect \\
rogat@kft.umcs.lublin.pl}

\date{\today}

\maketitle

\begin{abstract}
A sigma model formulation for static dilaton black holes with
$N$ Abelian gauge fields was presented. We proved uniqueness of
static asymptotically flat spacetimes with non-degenerate  black holes for
the considered theory.
\end{abstract}


\baselineskip=18pt
\section{Introduction}
Nowadays, much effort is being devoted to the study of mathematical topics
related to the black hole equilibrium states. The pioneering investigations 
in the subject was begun by Israel's researches \cite{isr67,isr68}.
It follows that Schwarzschild and Reissner-Nordstr\"om (RN) solutions are only
Einstein or Einstein-Maxwell (EM) (non-extreme) solutions that satisfy the 
conditions of being static  black hole metrics. Then,
M\"uller zum Hagen {\it et al.} \cite{mil73} and Robinson
\cite{rob77}  presented the generalization  of Israel's theorems.
\par
Different approach was proposed by Bunting and Masood-ul-Alam \cite{bun} 
who looked for conformal transformations on a spacelike hypersurface.
Next, the positive mass theorem \cite{ya,wi} 
enables them to draw conclusions that the considered hypersurface is
conformally flat.
Ruback \cite{ru} and
Masood-ul-Alam \cite{ma1} 
extended  this method to EM system, while 
Heusler \cite{he1} comprised the magnetically charged RN solution.
Chru\'sciel \cite{chr99a}
has finished the classification of static 
vacuum black holes. He removed the condition of non-degeneracy of the 
event horizon and showed that Schwarzschild black hole exhausted
the family of all appropriately regular black hole spacetimes. He also
revealed \cite{chr99b} that RN solution comprised the family of regular
black hole spacetimes under the restrictive condition that all
degenerate components of black hole horizon carried a charge of the same sign.
\par
The turning point for establishing the uniqueness of stationary
and axisymmetric black hole spacetimes being the solution of vacuum 
Einstein equations were achieved by Carter \cite{car73,car87}, 
Robinson \cite{rob75} and Wald
\cite{wal71}. The systematic way of obtaining the desire results in
electromagnetic case was conceived by Mazur 
\cite{maz82,maz84,maz84a,maz01} and Bunting \cite{bun}.
In Mazur's approach the key observation  was that the Ernst equations
depicted  a sigma model on a symmetric space, while Bunting's
approach was based on using a general class of harmonic mapping between
Riemaniann manifolds.\\
Considering $(n-1)$ Abelian gauge charged Kerr black hole G\"urses
\cite{gur84}
found that it was a unique stationary black hole solution of Einstein-Abelian 
gauge field equations. For a review of the uniqueness of black hole
solutions story see, for example \cite{maz01} and \cite{heub}.
\par
Nowadays, there has been an active period of constructing black hole solutions
in the string theories (see \cite{you97} and references therein). 
In \cite{mas93} Masood-ul-Alam
strengthened the method used in \cite{bun87} to the case of 
Einstein-Maxwell-dilaton (EMD) black holes. Using a sigma model formulation
of equations of motion for static dilaton black hole G\"urses {\it et al.}
\cite{gur95}
presented an alternative proof of the uniqueness of a static charged
dilaton black holes. Uniqueness theorems for black holes
are closely related to the problem of staticity. In the low-energy
limit of the heterotic string theory it was studied in \cite{rog97,rog98}.
In \cite{rog99} uniqueness theorem for static dilaton $U(1)^2$
black holes being the solution of $N = 4, d = 4$ supergravity
was presented. Black hole solutions were considered both in $SU(4)$
and $SO(4)$ versions of the underlying theory. Recently,
Mars {\it et al.} \cite{mar01}, using the 
{\it conformal positive mass theorem}
\cite{sim99} were able to extend the uniqueness proof of dilaton black holes.
The extension was valid for non-vanishing magnetic and electric charge, when
the coupling constant was $\alpha = 1$, or for arbitrary coupling constant 
but either vanishing magnetic or electric charge.
\par
Our paper is organized as follows. In section II
we consider static axially symmetric solution of EMD gravity with $N$-gauge
Abelian fields. 
Each of the gauge potential one-form is in the direction of time 
and in the azimuthal angle. By means of the introduced  pseudopotentials  
we arranged equations of motion in two complex Ernst-like relations. Using
the way of proving uniqueness of solutions for non-linear sigma-models
devised by Mazur \cite{maz82,maz84,maz01}
we established that two solutions of the boundary value problem were
equal if they were subject to the same set of boundary and
regularity conditions. As an example we showed the uniqueness
of Einstein-Maxwell-axion-dilaton
(EMAD) gravity black holes with axion fields being trivial.
In section III we concluded our results.

\section{Uniqueness Theorem}
We shall consider the effective action of the low-energy string theory,
Einstein-Maxwell-dilaton gravity which contains scalar field (dilaton)
$\phi$, Abelian gauge fields $A_{\mu}^{(i)}$,
where $i = 1, \ldots, N$
and the metric $g_{\mu \nu}$. For the sake of
generality we shall prefer to leave the arbitrary number of vector fields
(see, e.g., \cite{ber96,rog97,loz99}).
The action is of the form
\be
I = \int d^4 x \sqrt{-g} \bigg[ R - 2(\na \phi)^{2}
- e^{-2\phi} \sum\limits_{i = 1}^N
F_{\alpha \beta_(i)} F^{\alpha \beta (i)} 
\bigg],
\label{act}
\ee
where the strength of the $i-th$ gauge field is described by
$F_{\mu \nu}^{(i)} = 2\na_{[\mu} A^{(i)}_{\nu]}$.
The resulting equations of motion, derived from the variational
principle yield
\ben
R_{\mu \nu} = e^{-2 \phi} \sum\limits_{i = 1}^N
\bigg(
2 F_{\mu \rho (i)} F_{\nu}{}{}^{\rho (i)} - 
{1 \over 2} g_{\mu \nu}F^{(i) 2} \bigg) +
2 \na_{\mu} \phi \na_{\nu} \phi, \\
\na_{\mu} \na^{\mu} \phi + {1 \over 2} e^{-2\phi} 
\sum\limits_{i = 1}^N F^{(i) 2} = 0,\\ 
\na_{\mu} \bigg( e^{-2 \phi} F^{\mu \nu (i)} \bigg) = 0.
\een
We introduce the static axially symmetric line element
expressed as
\be
ds^2 = - e^{2 \psi} dt^2 + e^{-2 \psi} \left [
e^{2 \ga} \left ( dr^2 + dz^2 \right ) + r^2 d\phi^2 \right ],
\label{gij}
\ee
where $\psi $ and $\ga$ depended only on $r$ and $z$ coordinates.
The components of each gauge potential one-form $A^{(i)}$
will be in the direction of time and in the azimuthal
angle, namely
\be
A^{(i)} = A^{(i)}_{0} dt + A_{\phi}^{(i)} d\phi.
\ee
They were also assumed to depend only on
$r$ and $ z$ coordinates.
The resulting Eqs. of motion imply
\be
\na^2 \psi - e^{-2\psi - 2\phi} 
\sum\limits_{i = 1}^N
\bigg(
A_{0, r}^{(i) 2} + A_{0, z}^{(i) 2} \bigg) - {e^{2\psi - 2 \phi} \over r^2}
\sum\limits_{i = 1}^N \bigg( A^{(i)2}_{\phi, r} + A^{(i) 2}_{\phi, z}
\bigg) = 0,
\label{em1}
\ee
\be
\na^2 \phi - e^{-2\psi - 2\phi} 
\sum\limits_{i = 1}^N
\bigg(
A_{0, r}^{(i) 2} + A_{0, z}^{(i) 2} \bigg) + {e^{2\psi - 2 \phi} \over r^2}
\bigg(
 A^{(i)2}_{\phi, r} + A^{(i) 2}_{\phi, z}
\bigg) = 0,
\label{em2}
\ee
\be
\na_{r} \bigg( r e^{- 2 \psi - 2 \phi} A^{(i)}_{0, r}\bigg) +
\na_{z} \bigg( r e^{- 2 \psi - 2 \phi} A^{(i)}_{0, z} \bigg) = 0.
\label{em3}
\ee
\be
\na_{r} \bigg(  {e^{ 2 \psi - 2 \phi} \over r} A^{(i)}_{\phi, r} \bigg) +
\na_{z} \bigg(  {e^{ 2 \psi - 2 \phi} \over r} A^{(i)}_{\phi, z} \bigg) = 0.
\label{em4}
\ee
\be
{\ga_{, z} \over r} - 2 \psi_{, r} \psi_{, z} =
\sum\limits_{i = 1}^N \bigg(
- 2 e^{-2\psi - 2\phi} A^{(i)}_{0, r} A_{(i) 0, z} + 
{2 \over r^2} e^{2\psi - 2 \phi} 
A_{(i) \phi, r} A^{(i)}_{\phi, z} \bigg) + 2 \phi_{, r} \phi_{, z},
\label{em5}
\ee
\be
\sum\limits_{i = 1}^N \bigg(
e^{-2\psi - 2\phi} \left (
A^{(i) 2}_{0, r} - A^{(i)2}_{ 0, z } \right ) + 
{1 \over r^2} e^{2\psi - 2 \phi} 
\left ( A^{(i) 2}_{\phi, r} - A^{(i) 2}_{\phi, z} \right ) 
\bigg) +
\left ( \phi_{,z}^2 - \phi_{, r}^2 \right ) =
\psi_{, r}^2 - \psi_{, z}^2 - {\ga_{, r} \over r},
\label{em6}
\ee
Because of the fact that $\ga$ may be determined if we know
$\psi, \phi, A^{(i)}_{0}$ and $A^{(i)}_{\phi}$,
the essential part of the above system consists of equations (\ref{em1}-\ref{em4}).
Let us define the quantities
\be 
E = - \phi - \psi, \qquad B + \ln r = \psi - \phi,
\label{p1}
\ee
and
\be
\tiA^{(i)}_{0} = \sqrt{2} A^{(i)}_{0}, \qquad 
\tiA^{(i)}_{\phi} = \sqrt{2} A^{(i)}_{\phi}.
\label{p2}
\ee
Having in mind (\ref{p1}) and (\ref{p2}) we can combine relations
(\ref{em1}) and (\ref{em2}) as follows:
\be
\na^2 E + e^{2 E} \sum\limits_{i = 1}^N \na \tiA^{(i)}_{0} 
\na \tiA^{(i)}_{0}
= 0,
\label{ee}
\ee
\be
\na^2 B - e^{2 B} 
\sum\limits_{i = 1}^N \na \tiA^{(i)}_{\phi} \na \tiA^{(i)}_{\phi}
= 0,
\label{bb}
\ee
Inspection of (\ref{em3}) enables us to define a pseudopotential
$\om_{(i)}$
in  the form
\be
\om_{(i) r} = r e^{2 E} \tiA^{(i)}_{0, z}
, \qquad
\om_{(i) z} = - r e^{2 E} \tiA^{(i)}_{0, r}.
\label{w}
\ee
Our main aim is to rewrite relations (\ref{ee}-\ref{bb})
in the forms similar to the Ernst's equations. 
In order to do so  we introduce
the complex functions defined as:
\be
\ep_{1 (i)} = (r e^{E})_{(i)} + i \sqrt{N} \om_{(i)},
\ee
and
\be
\ep_{2 (i)} = (e^{- B})_{(i)} + i \sqrt{N} \tiA_{\phi (i)}.
\ee
Now equations (\ref{ee})-(\ref{bb}) and (\ref{em3})-(\ref{em4}) 
can be arranged 
into two complex, Ernst like relations
\be
\sum\limits_{i = 1}^N
\bigg( \ep_{1 (i)} + \bep_{1 (i)} \bigg) \na^2
\ep_{1}^{ (i)} = 2 \sum\limits_{i = 1}^N
\na \ep_{1 (i)} \na \ep_{1}^{(i)},
\label{e1}
\ee
\be
\sum\limits_{i = 1}^N
\bigg( \ep_{2 (i)} + \bep_{2 (i)} \bigg) \na^2
\ep_{2}^{ (i)} = 2 \sum\limits_{i = 1}^N
\na \ep_{2 (i)} \na \ep_{2}^{(i)}.
\label{e2}
\ee
Equations (\ref{e1}) and (\ref{e2}) can be obtained by the variation of the 
Lagrangian density
\be
{\cal L}_{m} = \sum\limits_{i = 1}^N {\na \ep_{m (i)} \na \bep_{m}^{(i)}
\over \big( \ep_{m (i)} + \bep_{m (i)} \big)^2},
\ee
where $m = 1,2$. We can change the formulation of the system by the 
homographic change of the variables we should have in mind that 
everything what will follow will be valid for both $\ep_{1 (i) }$
and $\ep_{2 (i)}$. Therefore from this stage on
for the simplicity of notation 
we shall drop the index $1$ and $2$. Consistently with this remark, we have
\be
\ep_{(i)} = {\xi_{i} - 1 \over \xi_{i} + 1},
\ee
where $i = 1,\ldots, N$. The Lagrangian density can be expressed as
\be
{\cal L} = \sum\limits_{i = 1}^N {\na \xi_{i} \na \bxi^{i}
\over \big( \xi_{i} \bxi^{i} - 1 \big)^2}.
\ee
The requirement of the positivity of 
$\big( \ep_{(i)} + \bep_{(i)} \big)^2$  implies that
$\xi_{i} \bxi^{i} - 1 > 0$,
and
the Lagrangian density defines the Bergmann metric $g_{\alpha {\bar \beta}}$
on the bounded symmetric domain in $C^N$. The metric becomes
\be
ds^2 ={1 \over \big( 1 - z_{\alpha} \bz^{\balpha} \big)^2}
dz_{\alpha}d \bz^{\balpha},
\ee
where $z_{\alpha} = \xi_{\alpha}$.
Defining new variables $\xi_{a} = v_{a}/v_{0}$, where $a = 1,\ldots, N$
we can easily rewrite $\cal L$ in the following form:
\be
{\cal L} = \sum\limits_{a = 1}^N { \mid
(\na v_{a}) v_{0} - (\na v_{0}) v_{a}
\mid^2 \over
\big( v_{a} \bv^{a}  - v_{0} \bv^{0} \bigg)^2}
\ee
The above Lagrangian density takes the manifestly $SU(N,1)$ invariant
form described as follows:
\be
{\cal L} = - {
\bigg[ \bv^{\alpha} v_{\alpha} (\na \bv^{\beta} \na v_{\beta})
- \bv^{\alpha} \na v_{\alpha} (\na \bv^{\alpha} v_{\alpha}) \bigg]
\over \big( \bv^{\alpha} v_{\alpha} \big)^2},
\ee
where $\alpha = 0, 1,\ldots, N$, $\eta_{\alpha \beta} = diag (1,\ldots,-1)$.
A complex vector $v_{\alpha}$ determines the projector \cite{maz82}
\be
P_{\alpha}{}{}^{\beta} = {v_{\alpha} \bv^{\beta} 
\over \bv^{\alpha} v_{\alpha}},
\ee
possessing the following properties:
\be
Tr P = {\bf 1}, \qquad P^2 = P, \qquad \bigg( \eta_{\alpha \beta} 
P^{\alpha \beta} \bigg)^{\dagger}  = \eta_{\alpha \beta} 
P^{\alpha \beta},
\ee
where $\bf 1$ is the unit matrix. 
Definition of $P_{\alpha}{}{}^{\beta}$ enables one to prove the identity
\be
{1 \over 2} \bigg( \na P_{\alpha \beta} \na P^{\alpha \beta} \bigg) =
{1 \over (\bv^{\alpha} v_{\alpha})^2}
\bigg[ \bv^{\alpha} v_{\alpha} \na \bv^{\ga} \na v_{\ga} -
\bv^{\delta} \na v_{\delta} \na \bv^{\beta} v_{\beta} \bigg].
\ee
Let us introduce next a matrix $\Phi = {\bf 1} - 2 P$.
One can verify the following relation:
\be
{1 \over 2} Tr \big( \na \Phi \na \Phi^{-1} \big)
 = 2 Tr (\na P \na P).
\ee
The matrix $\Phi$ is positive and hermitian.
Defining the quantity $j_{\mu} = \na_{\mu}\Phi \Phi^{-1}$ one can see that the 
Lagrangian density may be written as ${\cal L} = - {1/8}j_{\mu} j^{\mu}$.
The Eqs. of motion for sigma-model are of the form
\be
\na_{\mu} j^{\mu} = 0.
\ee
\par
Using the definition of $\xi_{\alpha}$ one can write $\Phi_{\alpha
\beta}$ in the Ernst's parameterization as follows:
\be
\Phi_{\alpha \beta} = \eta_{\alpha \beta} - {2 \xi_{\alpha} \bxi_{\beta}
\over \bxi^{\alpha} \xi_{\alpha}}.
\ee
In particular for $\Phi_{\alpha \beta}$  we have
\be
\Phi_{\alpha \beta} = {1 \over  \sum\limits_{i = 1}^N \xi_{i} \bxi^{i}}
\pmatrix{ 1 + \sum\limits_{i = 1}^N
\xi_{i} \bxi^{i}  & 2 \bxi_{1} & 2 \bxi_{2} &
\ldots& 2\bxi_{N}
\cr 2 \xi_{1} & 1+ \xi_{1} \bxi_{1} - \sum\limits_{i = 2}^N
\xi_{i} \bxi^{i} & 2 \xi_{1} \bxi_{2}&\ldots& 2 \xi_{1} \bxi_{N} \cr
\vdots& \vdots& \vdots & \vdots& \vdots \cr
2 \xi_{N}& 2 \xi_{N} \bxi_{1}&\ldots & \ldots&
1 - \sum\limits_{i = 1}^{N-1} \xi_{i} \bxi^{i} +
\xi_{N} \bxi^{N}}.
\ee
The most difficult part in proving the uniqueness theorem is to show
that two solutions of the boundary value problem are equal if they are
subject to the same set of boundary and regularity conditions. 
We shall follow the method used for the first time in 
\cite{maz82} to prove the above.
To establish the uniqueness of solutions of the
Ernst equations (\ref{e1}) and (\ref{e2}) we have to calculate
the exact form of $Tr ( \Phi_{1} \Phi_{2}{}{}^{-1})$ in
terms of $E_{1,2}, \omega_{(i)1,2}$ and $B_{1,2}, \tiA_{\phi (i)1,2}$. After
a straightforward calculations one finds, respectively,
\ben \label{tr1}
q(E, \omega_{(i)}) &=&
Tr ( \Phi(E, \omega)_{1} \Phi(E, \omega)_{2}{}{}^{-1}) = \eta_{\alpha}{}{}^{\alpha} +
{1 \over r^2 e^{E_{1} + E_{2}}} \bigg[
\big( \omega_{(i)1} - \omega_{(i)2} \big)
\big( \omega^{(i)}_{1} - \omega^{(i)}_{2} \big)
+ r^2 \big( e^{E_{1}} - e^{E_{2}} \big)^2 \bigg],\\
q(B, \tiA_{\phi (i)}) &=&
Tr ( \Phi(B, \tiA_{\phi})_{1} \Phi(B, \tiA_{\phi})_{ 2}{}{}^{-1}) = \\ \nonumber
&=& \eta_{\alpha}{}{}^{\alpha} 
+ {1 \over e^{- (B_{1} + B_{2})}} \bigg[
\big( \tiA_{\phi (i) 1} - \tiA_{\phi (i) 2} \big)
\big( \tiA^{(i)}_{\phi~ 1} - \tiA^{(i)}_{\phi~ 2} \big)
+ \big( e^{- B_{1}} - e^{- B_{2}} \big)^2 \bigg].
\label{tr2}
\een
Let us now proceed to the uniqueness proof.\\
To prove the uniqueness let us
consider two solutions to the $\sigma$-model 
equations, i.e., $\Phi_{1}$
and $\Phi_{2}$ being subject to the same boundary conditions.
Next, define $\Phi = \Phi_{1} \Phi_{2}^{- 1}$ and evaluate 
the Laplacian, and take the trace of this relation. One obtains
a relation of the form
\ben \label{tr}
Tr \bigg( \na_{\mu} \na^{\mu} \Phi \bigg) +
Tr \bigg[ \Phi  \bigg( \na_{\mu}( j^{\mu (2)}) - \na_{
\mu}( j^{\mu (1) } \bigg) \bigg] = \\ \nonumber
Tr \bigg[
\Phi \bigg( j_{\alpha}^{(1)}j^{\alpha (1)} +
j_{\alpha}^{(2)}j^{\alpha (2)} - 2 j_{\alpha}^{(2)}j^{\alpha (1)}
\bigg) \bigg],
\een
where $j_{\alpha}^{(i)} = \na_{\alpha} \Phi^{(i)}\Phi^{(i) -1}$. 
Having in mind
the fact that $\Phi$ is a Hermitian positive definite matrix
we can 
decompose $\Phi = g g^{\dagger}$, where $g \in SU(N,1)$,
one can find that the right-hand side 
of (\ref{tr})
can be rearrange in the form
$RHS = Tr (A A^{\dagger}) \ge 0$,
where $A = g_{2}^{-1} (j^{(1)} - j^{(2)})g_{1}$ \cite{maz82}.
\par
Applying Stoke's theorem
to Eq.(\ref{tr}) one obtains
\be
\int_{S}dS~ Tr \bigg[ \Phi \big(
\na_{\mu} j^{\mu (2)} -  \na_{\mu} j^{\mu (1)} \big)  \bigg] +
\int_{\p S} d S^{\mu} \na_{\mu} \big( Tr \Phi \big) =
\int_{S} dS~ Tr \big( A_{\mu} A^{\dagger \mu} \big) \ge 0,
\label{trr}
\ee
where $dS^{\mu}$ denotes the components of the 
metric normal surface element.\\
The above identity serves to complete the proof of uniqueness
theorem. The first term on the left-hand side of (\ref{trr})
vanishes identically because of
fulfilling equations of motion on the boundary $\p S$.
The second term vanishes if the boundary conditions are taken into
account. On the other hand, the right-hand side integral is non-negative.
Since the left-hand side integral over the region $S$ vanishes, then
the right-hand side of relation (\ref{trr}) must vanish identically on the
solutions $\Phi_{1}$ and $\Phi_{2}$. Thus vanishing of the 
right-hand side implies 
that $Tr \Phi = const$ and $\Phi_{1} = \Phi_{2}$. In our case from 
(\ref{tr1}) and (\ref{tr2}) $const = \eta_{\alpha}{}{}^{\alpha} = N$.
We arrive at the result:\\
\noindent
{\bf Theorem:}\\
Let $\Phi_{1}$ and $\Phi_{2}$ be the solution of sigma model
equations on $SU(N,1)/S(U(N) \times U(1))$ with 
two-dimensional base manifold satisfying the
equations of motion for sigma model
$\na_{\mu} j^{\mu} = 0$.
Moreover, let $\na_{\mu}( Tr \Phi ) = 0$, where 
$\Phi = \Phi_{1} \Phi_{2}{}{}^{-1}$
on the boundary $\p S$ of the region $S$. Then, $\Phi_{1} = \Phi_{2}$
in all region $S$ provided that $\Phi_{1}(p) = \Phi_{2}(p)$
for at least one point $p \in S$.

\vspace{0.1cm}
In \cite{rog99} we considered black hole solutions in which each gauge field 
has either electric or magnetic charge. The above proof enables us to
generalize the uniqueness of black solutions to solutions  in which both 
gauge fields have electric and magnetic charge (this is $N = 2$ case in 
the present derivation). 
As an example one can consider black holes appearing in the doubly 
charged electro-magnetic
solution of EMAD gravity with axion field being trivial \cite{ort93}.
The effective action for this theory is as follows:
\ben \label{act1}
I_{EMAD} =
\int d^4 x \sqrt{-g} \bigg[ R - 2(\na \phi)^{2}
-{1 \over 2} e^{4 \phi} (\na a)^2
&-& e^{-2\phi} \big(
F_{\alpha \beta} F^{\alpha \beta } +
G_{\alpha \beta} G^{\alpha \beta } \big)  \\ \nonumber
&+& a \big( F_{\mu \nu} \ast F^{\mu \nu} +
G_{\mu \nu} \ast G^{\mu \nu} \big)
\bigg],
\een
where $F_{\alpha \beta}$ and $G_{\alpha \beta}$ are two $U(1)$ fields and
the dual of an antisymmetric tensor is defined as
$\ast F_{\mu \nu} = {1 \over 2} \ep_{\mu \nu \alpha \beta} F^{\alpha
\beta}$ for $F_{\alpha \beta}$ and in the same manner for $G_{\alpha \beta}$
.\\
Equations of motion yield
\ben
\na^2 \phi - {1 \over 2} e^{4 \phi}(\na a)^2 + {1 \over 2} e^{- 2 \phi}
\big( F^2 + G^2 \big) = 0, \\ 
\na_{\mu} \ast F^{\mu \nu} = 0, \qquad \na_{\mu} \ast G^{\mu \nu} = 0,\\
\na_{\mu} \big(
e^{- 2 \phi} F^{\mu \nu} + a \ast F^{\mu \nu} \big) = 0,\\
\na_{\mu} \big(
e^{- 2 \phi} G^{\mu \nu} + a \ast G^{\mu \nu} \big) = 0,\\
\na_{\mu} \big( e^{4 \phi} \na^{\mu} a \big) - \big(
F_{\mu \nu} \ast F^{\mu \nu} + G_{\mu \nu} \ast G^{\mu \nu} \big) = 0,\\
R_{\mu \nu} = e^{- 2 \phi}
\bigg[ \big(
2 F_{\mu \rho}F_{\nu}{}{}^{\rho} - {1 \over 2}g_{\mu \nu} F^2 \big)
+
\big(
2 G_{\mu \rho}G_{\nu}{}{}^{\rho} - {1 \over 2}g_{\mu \nu} G^2 \big)
\bigg]
+ 2 \na_{\mu} \phi \na_{\nu} \phi + {1 \over 2} e^{4 \phi} \na_{\mu }a \na_{\nu} a.
\een
Having in mind equations of motion derived from
(\ref{act1}) we see that this is the case 
of two gauge one-form potentials $A^{(1)}$ and $A^{(2)}$ corresponds
to the case when
axion field is equal to zero.
We remark that
our proof comprises also the case studied by Ortin \cite{ort93},
when axion field is trivial (being constant). But in this case the auxiliary
condition appears, namely $F_{\mu \nu} \ast F^{\mu \nu} +
G_{\mu \nu} \ast G^{\mu \nu} = 0$.

\section{Conclusions}
In our work we have studied EMD gravity with $N$ gauge fields, being the
generalization  of the simple EMD gravity with one gauge field
widely studied in literature. We were interested in uniqueness theorem
for black hole solutions in the considered theory. It was assumed that
the gauge potential one-form $A^{(i)}$ has two components, in the direction
of time and the azimuthal angle. Finding the adequate pseudopotentials
we managed to assemble the equations of motion in two Ernst's like relations.
Using the idea of Mazur \cite{maz82} that these equations described a non-linear
sigma model on the symmetric space we established the uniqueness of
black hole solutions.
\par
In \cite{rog99} we considered black hole solutions where the gauge 
fields have either electric or magnetic charge. A natural generalization of
these solution is to consider the case when the gauge fields have both
electric and magnetic charges. The present proof enables one to 
treat this case and to find the uniqueness of such solutions      .
\par
On the other hand, when we consider the doubly charged black hole solutions
in EMAD gravity with axion field being trivial \cite{ort93}, one should fulfill
the auxiliary condition of the form  $F_{\mu \nu} \ast F^{\mu \nu} +
G_{\mu \nu} \ast G^{\mu \nu} = 0$. Then, for $N = 2$ one has the uniqueness
of the black hole solution in this theory.
Of course, the challenge is to prove the uniqueness theorem when axion fields
are arbitrary. We hope to return to this problem elsewhere.

\vspace{1cm}
\noindent
{\bf Acknowledgements:}\\
I would like to thank the unknown referees for useful comments.\\
MR was supported in part by KBN grant 5 P03B 009 21.



\end{document}